 \documentclass[final,5p,times,amsmat,twocolumn]{elsarticle}

 \usepackage{graphics}

\usepackage{amssymb}
 \usepackage{amsthm}

\newcommand{\beq}{\begin{equation}}  
\newcommand{\eeq}{\end{equation}}  
\newcommand{\beqa}{\begin{eqnarray}}  
\newcommand{\eeqa}{\end{eqnarray}}

\usepackage{xcolor}

\journal{Surface Science}

\begin{document}

\begin{frontmatter}



\title{Consequences of Kondo exchange on quantum spins}





\author[inl]{F. Delgado\corref{cor1}}
\ead{fernando.delgado@inl.int}

\author[ucl,cucl,pucl]{C. F. Hirjibehedin}

\author[inl]{J. Fern\'andez-Rossier\fnref{fn2}} 


\cortext[cor1]{Corresponding author at International Iberian Nanotechnology Laboratory (INL), 4715-330 Braga, Portugal. Tel: +351 253 140 112; fax: +351 253 140 119. }

\fntext[fn2]{On leave from Departamento de F\'{i}sica Aplicada, Universidad de Alicante, Spain }

\address[inl]{International Iberian Nanotechnology Laboratory (INL), 4715-330 Braga, Portugal.}

\address[ucl]{London Centre for Nanotechnology, University College London (UCL), London WC1H 0AH, UK}

\address[cucl]{Department of Physics and Astronomy, UCL, London WC1E 6BT, UK.}
\address[pucl]{Department of Chemistry, UCL, London WC1H 0AJ, UK}

\begin{abstract}
When individual quantum spins are placed in close proximity to conducting substrates, the localized spin is coupled to the nearby itinerant conduction electrons via Kondo exchange. In the strong coupling limit this can result in the Kondo effect - the formation of a correlated, many body singlet state - and a resulting renormalization of the density of states near the Fermi energy. However, even when Kondo screening does not occur, Kondo exchange can give rise to a wide variety of other phenomena. In addition to the well known renormalization of the $g$ factor and the finite spin decoherence and relaxation times, Kondo exchange has recently been found to give rise to a newly discovered effect: the renormalization of the single ion magnetic anisotropy. Here we put these apparently different phenomena on equal footing by treating the effect of Kondo exchange perturbatively. In this formalism, the central quantity is $\rho J$, the product of the density of states at the Fermi energy $\rho$ and the Kondo exchange constant $J$. We show that perturbation theory correctly describes the experimentally observed exchange induced shifts of the single spin excitation energies, demonstrating that Kondo exchange can be used to tune the effective magnetic anisotropy of a single spin.

\end{abstract}


\end{frontmatter}


\section{Introduction\label{intro}}
The development of electron paramagnetic resonance made it possible to study the spin transitions of a variety of spin systems, such as paramagnetic molecules \cite{Tinkham_PhD_1954} and transition metal dopants in insulating hosts~\cite{Abragam_Bleaney_book_1970}. This led to the development of a single spin Hamiltonian, where the influence of both the Zeeman effect and magnetic anisotropy determine the energy spectrum and spin selection rules. Interestingly, the same type of Hamiltonian was successfully used to describe the quantum spin tunnelling phenomenon \cite{Gatteschi_Sessoli_book_2006} discovered in magnetic molecules with large spin. 

Thanks to the tremendous progress in nano-fabrication and nano-manipulation, it is now possible to produce devices where an individual quantum spin can be probed.  A single magnetic molecule can be placed in a nanoscale junction \cite{Scott_ACSNano_2010,Zyazin_Nanolett_2010,Parks_Science_2010,Vincent_Klyatskaya_nat_2012}, on top of a carbon nanotube \cite{Ganzhorn_Klyatskaya_acs_2013}, or on a surface \cite{Mannini_Nature_2010,Kahle_Deng_nanolett_2011}. A particularly suitable instrument for studying spin systems at the atomic scale is a scanning tunnelling microscope (STM) because it permits not only probing but also manipulation of the spin of individual magnetic atoms deposited on surfaces \cite{Wiesendanger_revmod_2009,Brune_Gambardella_sursci_2009,Gauyacq_Pierre_psc_2012}, which thereby takes us closer to the Feynman's dream of engineering matter at the atomic scale. Interestingly, magnetic adatoms can also be described with the same type of single ion Hamiltonian as magnetic dopants in insulating hosts and single-molecule magnets \cite{Hirjibehedin_Lin_Science_2007}. These systems have attracted significant attention because they represent the ultimate limit of magnetic objects where classical or quantum information can be stored \cite{Leuenberger_Loss_nature_2001,Delgado_Rossier_prl_2012}.

Manipulation and readout of the information requires integration of these quantized spins (e.g. magnetic molecules, spin chains, or magnetic atoms) into a device. In the case of quantum spins in contact with a conducting electrode, found in many proposed device geometries \cite{Bogani_Wernsdorfer_natmat_2008}, the question of how exchange interaction with the conduction electrons changes the spin dynamics of the quantized spin naturally arises. In the strong coupling regime, the Kondo effect is known to quench the magnetic moment of the quantum spin. This comes together with a strong renormalization of the states at the Fermi energy, which in STM measurements is revealed as a Fano lineshape in the low bias conductance \cite{Madhavan_Chen_1998,Ternes_JPCM_2009}.

\begin{figure}[t]
  \begin{center}
  \includegraphics[width=1.\linewidth,angle=0]{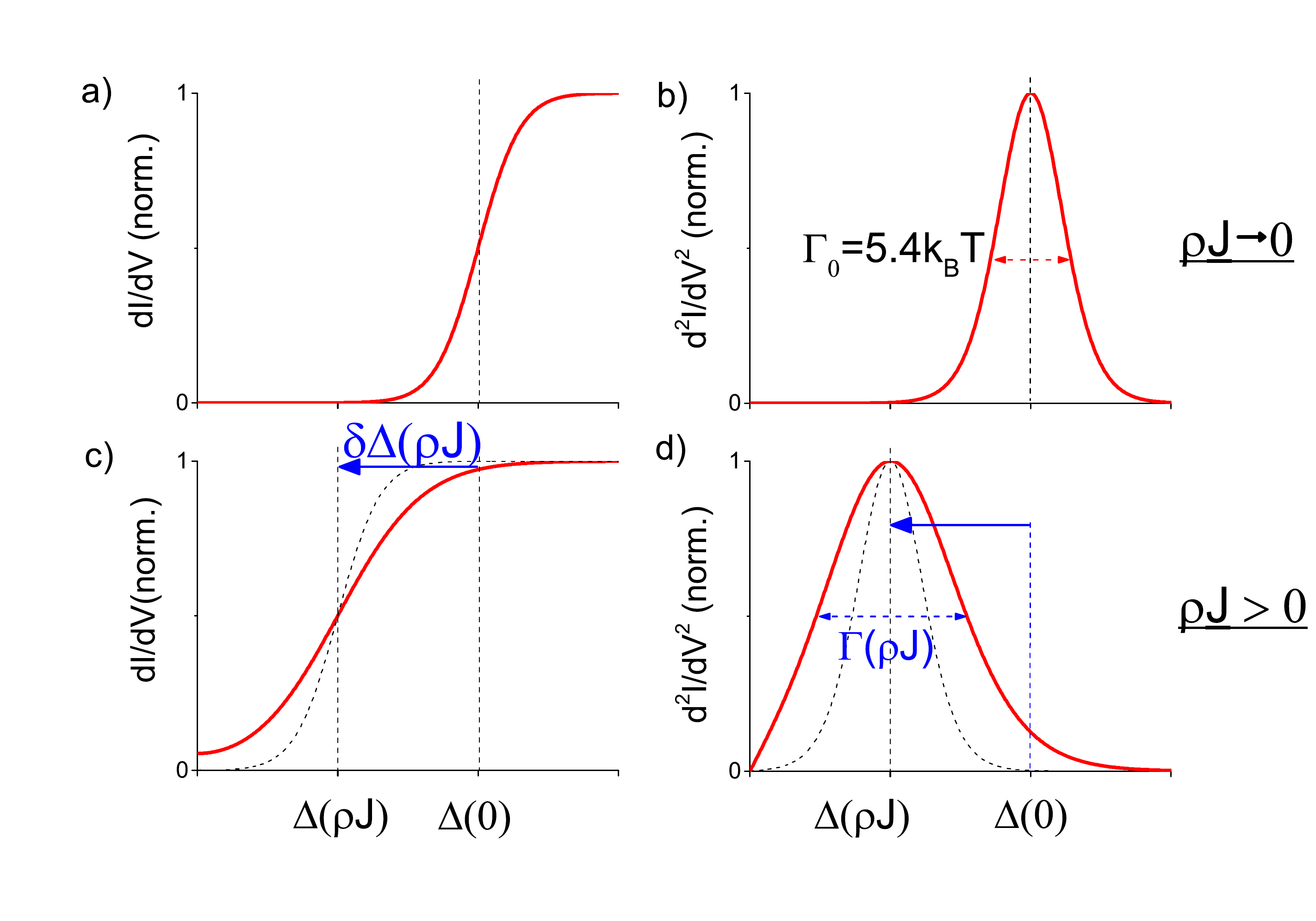} 
  \end{center}
  \caption{Main effects of the Kondo coupling on the electronic transport. The inelastic spin-flip transitions are revealed as thermally broadened steps in the $dI/dV$ [panels a) and c)], and peaks (or dips for negative voltage) in the $d^2I/dV^2$, panels b) and d). As the Kondo coupling grows ($\rho J$ increases, lower panels), the inelastic step shifts to lower energies and it broadens. The black-dotted lines 
in panels c) and d) represent the shapes of the $dI/dV$ and $d^2I/dV^2$ respectively considering only the shift $\delta\Delta(\rho J)$ in the transition energy and neglecting the change in the broadenings $\hbar \Gamma_{n\to m}^{SS}$, Eq. (\ref{ratesg}), while the new shapes appear in red. 
%
    \label{fig1}
  }
\end{figure}

More recently, the tunnelling spectra of magnetic adatoms \cite{Heinrich_Gupta_science_2004,Hirjibehedin_Lutz_Science_2006,Hirjibehedin_Lin_Science_2007,Otte_Ternes_natphys_2008,Loth_Bergmann_natphys_2010,Khajetoorians_Chilian_nature_2010,Bryant_Spinelli_prl_2013,Khajetoorians_Schlenk_prl_2013,Miyamachi_Schuh_nature_2013} and molecules \cite{Chen_Fu_prl_2008,Tsukahara_Noto_prl_2009} have been found to display inelastic spin transitions,  revealed as magnetic field-dependent steps in the differential conductance $dI/dV$ (see Fig. \ref{fig1}).  Fitting the energies of these steps to an effective spin Hamiltonian provides a quantitative understanding of the magnetic anisotropy \cite{Hirjibehedin_Lutz_Science_2006,Hirjibehedin_Lin_Science_2007,Delgado_Rossier_prb_2010,Khajetoorians_Chilian_nature_2010,Miyamachi_Schuh_nature_2013}. The steps in $dI/dV$ are equivalently peaks in $d^2I/dV^2$, whose half width at half maximum comes from thermal and instrumental smearing as well as the broadening of the transition due to the finite spin lifetime. 
  
Importantly, Kondo exchange influences the quantum spins even in the absence of Kondo effect, i.e. when no Kondo feature is seen in the conductance spectrum.   For instance,  because of the Kondo exchange, the single-spin states acquire a finite lifetime \cite{Langreth_Wilkins_prb_1972,Delgado_Rossier_prb_2010}. In the case of magnetic adatoms, fast spin relaxation times, of the order of 200 fs, have been estimated from the full width at half maximum of the $d^2 I/dV^2$ peaks of Fe atoms on a metal \cite{Khajetoorians_Lounis_prl_2011}, while direct STM measurements of the relaxation times of Fe on top of a Cu$_2$N substrate, possible using pump and probe techniques \cite{Loth_Etzkorn_science_2010}, has demonstrated lifetimes up to 50 ns.

The Kondo exchange can actually arise from  two different physical mechanisms. First, direct ferromagnetic exchange is possible between the itinerant electrons of the surface and the $d$ or $f$ levels of the atomic spin.  This type of exchange is responsible, for instance, for the spin splitting of the conduction $s$-type band in diluted magnetic semiconductors~\cite{Furdyna_jap_1988}.  Second, if the surface electrons hybridize with the localized $d$ or $f$ orbitals,  the so called kinetic exchange~\cite{Anderson_prl_1966}  results in an antiferromagnetic Kondo coupling, in the limit where classical charge fluctuations of the atom are frozen. This second mechanism is almost ubiquitous and can coexist with the first, giving rise to a reduced total exchange due to their opposite signs. 


In this work, we emphasize the central role of the Kondo exchange coupling in a vast variety of available experimental observations of magnetic adsorbates, 
and thus, we demonstrate that it is possible to quantify its effects. Table \ref{table} shows a summary of physical quantities associated to the spins of a few magnetic impurities that are modified by the exchange coupling to the conduction electrons. Notice that all of them depends on the product of the electrode density of states at the Fermi level, $\rho$, and the Kondo exchange coupling $J$. The effect can be classified then according to the order in $\rho J$. To first order it leads to a modification of the effective g factor, an effect akin to the  Knight-shift in metals. 
To second order, it leads to finite decoherence and lifetimes~\cite{Delgado_Rossier_prb_2010,Delgado_Rossier_prl_2012,Khajetoorians_Lounis_prl_2011,Loth_Etzkorn_science_2010} or the indirect exchange due to the RKKY interaction~\cite{Kittel_book_1963}.
In addition to these well known results, the Kondo exchange coupling also leads to another second order effect recently observed in  magnetic atoms: the renormalization of the magnetic anisotropy~\cite{Oberg_Calvo_natnano_2013}. 
Perturbation theory breaks down either when $\rho J$ is not a small parameter, in the case of ferromagnetic $J$, or  below the Kondo temperature \cite{Hewson_book_1997},  $k_B T_K = W e^{-1/(\rho J)}$, in the case of antiferromagnetic $J$.

\begin{table}
\centering
\begin{tabular}{lccc}
\hline
\hline
{\bf Quantity} & {\bf Symbol}  & {\bf Equation} &   {\bf Reference} \\
\hline
 g factor & $g^* $          &  $g(1-\frac{g_e}{2 g_A}\rho J)$  &  \cite{Wolf_Loose_physlett_1969} \\
 Spin relaxation &$\hbar T_1^{-1} $          &  $(\rho J)^2  S^2 \Delta$  &  \cite{Delgado_Rossier_prb_2010} \\
 Spin decoherence  & $\hbar T_2^{-1} $          &  $(\rho J)^2 S^2 k_B T$  &  \cite{Delgado_Rossier_prl_2012} \\
 Indirect exchange  & $J_{RKKY}^{-1} $          &  $(\rho J)^2$ F(r)  &  \cite{Kittel_book_1963} \\
Exchange shift    & $\delta \Delta$          &  $\propto (\rho J)^2 \ln\frac{2W}{\pi k_B T}$    &  \cite{Oberg_Calvo_natnano_2013} \\
Kondo Temperature & $k_B T_K $          &  $ We^{-1/(\rho J)^2}$  &  \cite{Hewson_book_1997} \\
 \hline
 \end{tabular}
 \label{table}
 \caption{Physical quantities associated with the Kondo exchange coupling $J$ between a magnetic impurity and conduction electrons spins. All of them are determined by the product $\rho J$, with $\rho$ the substrate density of states at the Fermi level. }
\end{table}

%

\section{Theoretical approach}
\subsection{Hamiltonian model\label{hamil}}

Our starting point is the Hamiltonian~\cite{Delgado_Rossier_prb_2010,Rossier_prl_2009,Fransson_nanolett_2009}
\begin{equation}
  {\cal H}=  {\cal H}_{\rm S} + {\cal H}_{\rm surf} + {\cal V}_{\rm kondo},
  \label{:sd-model}
\end{equation}
where ${\cal H}_{\rm S}$ is a single spin Hamiltonian discussed below, and ${\cal H}_{\rm surf}$ describes the independent electrons of the surface
\begin{equation}
  {\cal H}_{\rm surf}=\sum_{k,\sigma}\epsilon_{k,\sigma} c^{\dagger}_{k\sigma}c_{k\sigma},
  \label{surf}
\end{equation}
with $c_{k\sigma}^\dag$ ($c_{k\sigma}$) the creation (annihilation) operator of an electron 
in the surface with momentum $k$, spin $\sigma$ and single particle energy $\epsilon_{k\sigma}$.
Except in Sec. \ref{linearm},  we consider non-magnetic surfaces where $\epsilon_{k,\sigma}=\epsilon_{k}$.
 Finally,  
${\cal V}_{\rm kondo}$ describes the local exchange interaction between the surface electron density and the magnetic adatoms:
\begin{equation}
{\cal V}_{\rm kondo}= \frac{1}{2}\sum_{\vec k,\vec k',\sigma,\sigma'} J_{\vec{k},\vec{k}'}
 \vec{S}\cdot\vec{\tau}_{\sigma,\sigma'}
 c^{\dagger}_{k\sigma}c_{k'\sigma'},
 \label{Kondo}
\end{equation}
with $\vec{\tau}$ the vector of Pauli matrices (with $\pm1$ eigenvalues) and $J_{k,k'}$
   the s-d exchange interaction 
between the local spin $\vec S$ and the transport electrons. For simplicity, we will assume that these coupling constants are momentum independent and we shall take the average value on the Fermi surface, i.e., $J_{k,k'}=J/L^d$, where $L$ is a typical length of the macroscopic electrode and $d$ its dimensionality (i.e. $L^d$ can be either an area or a volume).
Notice that $J$ has dimensions of energy times $L^d$.  
 
In order to keep the ensuing  discussion as general as possible, we consider a  generic spin Hamiltonian
in Eq. (\ref{:sd-model}), 
${\cal H}_S  =\sum_n E_n |n\rangle\langle n|$
where $|n\rangle$ may describe both the electronic $\vec{S}$ and nuclear  $\vec{I}$ spin degree(s) of freedom.  For specific calculations we will use a Hamiltonian valid for 
 many magnetic systems on surfaces, including Mn, Co and Fe on Cu$_2$N/Cu~\cite{Hirjibehedin_Lin_Science_2007,Otte_Ternes_natphys_2008},  Fe-Pc molecules\cite{Tsukahara_Noto_prl_2009} and Fe clusters~\cite{Khajetoorians_Baxevanis_science_2013}
 \begin{equation}
  {\cal H}_{S} = D S_z^2 + E(S_x^2-S_y^2)  + g_A \mu_B \vec{B}\cdot\vec{S},
  \label{spin-hamil}
\end{equation}
where the first and second terms correspond to the axial and transverse magnetocrystalline anisotropy respectively and
$g_A$ stands for the adatom gyromagnetic factor. 
This Hamiltonian, which neglects the hyperfine coupling, is adequate for a system with 3 nonequivalent magnetic symmetry axis~\cite{Hirjibehedin_Lin_Science_2007}.
  On a surface the $D S_z^2$ term will be most certainly present, but other possibilities involving $S_x$ and $S_y$  are also possible~\cite{Miyamachi_Schuh_nature_2013}.
In the following we show the influence of the exchange interaction (\ref{Kondo}) on the dynamics of the anisotropic spin governed by a generic spin Hamiltonian ${\cal H}_S$.

\subsection{Calculation methodology\label{pertu}}
In this article we concentrate on the weak coupling regime where a perturbative treatment of the exchange interaction (\ref{Kondo}) is valid. This means that the dimensionless constant that controls the strength of the exchange coupling,  $\rho J$ as described below, with $\rho$ the volumetric density of states of the conduction electrons at the Fermi level, satisfies  $\rho J\ll 1$.

We now review the perturbative theory that permits addressing the influence of the Kondo exchange on the spin dynamics and the energy levels of the spin.  The main idea is that we treat the anisotropic spin as an open quantum system coupled, via Kondo exchange, to an electronic {\em reservoir}, given by the conduction electrons.   
In principle both the surface and tip electrons affect the atomic spin. The effect of the tip can be made negligible by decreasing the tunnelling amplitude.  Hence, we focus on the effect of the surface electrons, although the generalization to include the tip is straightforward \cite{Delgado_Rossier_prb_2010}.

\section{First order perturbation theory\label{linearm}}

To linear order in the Kondo exchange, the electron gas can influence the spin of the magnetic adatom provided that time reversal symmetry is broken, either spontaneously or by an applied field. When that happens, the average magnetization of the electron gas is nonzero.  Quantizing the spin along that axis, which we call $z$,  and replacing $c^{\dagger}_{k\sigma}c_{k'\sigma'}$ by its average $\langle c^{\dagger}_{k\sigma}c_{k'\sigma'}\rangle=
\delta_{k,k'}\delta_{\sigma,\sigma'} f(\epsilon_{k\sigma}) $, where $f(\epsilon)$ is the Fermi Dirac distribution, one gets that for a momentum independent coupling constant $J_{kk'}=J/L^d$, the Kondo coupling of Eq. (\ref{Kondo}) reads
as
\begin{equation}
 {\cal V}_{\rm kondo}\simeq S_z\sum_{k}
\frac{ J }{2L^d}\left[f(\epsilon_{k\uparrow})- f(\epsilon_{k\downarrow})\right].
 \label{Kondo-average}
\end{equation}
Then, if we introduce the total number of electrons with spin $\sigma$, $N_\sigma$, we get 
\begin{equation}
 {\cal V}_{\rm kondo}\simeq J S_z \frac{N_{\uparrow}-N_{\downarrow}}{2L^d}.
 \label{Kondo-average2}
\end{equation}
Thus, at this rough level of approximation, the spin of the magnetic adatom is coupled to the average spin density of the electron gas.   In the presence of an applied magnetic field,  simple linear response theory yields the well known
 Pauli paramagnetism result~\cite{Ashcroft_Mermin_book_1976}
\begin{equation}
\frac{N_{\uparrow}-N_{\downarrow}}{2L^d} = -\frac{M}{\mu_B g_e} = -\frac{\mu_Bg_e}{2}B \rho 
\end{equation}
where 
$g_e$ is the gyromagnetic factor of the surface electrons. 
We can now re-write the Zeeman term as $g^*\mu_B B S_z$ with the renormalized Land\'e $g^*$-factor 
\begin{equation}
g^*=g_A\left( 1 -\frac{g_e}{2g_A} \rho J \right)
\label{Knight}
\end{equation}
Thus, to linear order,  the effect of  the Kondo exchange is to renormalize the $g$ factor~\cite{Wolf_Loose_physlett_1969}, an effect that can only be measured in the presence of an applied magnetic field and was     first proposed in the context of nuclear magnetic resonance shifts in metals,  where it is known as Knight shift~\cite{Knight_pr_1949}. This $g$-factor renormalization has been recently detected by STM measurements~\cite{Zhang_Kahle_natcomm_2013}.

\section{Second order perturbation theory}

%
To first order in the Kondo coupling, the quantized spin couples only to the average magnetization.  To second order, in contrast, the quantized spin couples to the quantum fluctuations of the electron gas magnetization, which occur through the creation of electron-hole pairs across the Fermi energy. To second order in $\rho J$, this coupling has three main consequences on the quantized spin:
\begin{enumerate}
\item The eigenstates $|n\rangle$ of the quantized spin  acquire a finite lifetime $T_1(n)$.
\item If the quantized spin is prepared in 
a coherent superposition of states $n$ and $m$, 
it decays in a time scale $T_2(n,m)$.
\item The transition energy $\Delta_{nm}=E_n-E_m$ is renormalized due to the Kondo exchange .
\end{enumerate}
Whereas the notions of spin relaxation and spin decoherence are well established theoretically and experimentally, and have been worked  out in the context of exchanged coupled quantized spins \cite{Delgado_Rossier_prb_2010},  the notion of a reservoir induced shift is less common and  it has been only very recently reported in the context of magnetic adatoms \cite{Oberg_Calvo_natnano_2013}.

\subsection{Bloch-Redfield quantum master equation}

We now show that the second order perturbation theory is able to account for these three physical phenomena.  Formally, this is done using the so called system plus reservoir Bloch-Redfield \cite{Cohen_Grynberg_book_1998} approach, in which the  dissipative contribution  of a {\it reservoir} on the otherwise coherent dynamic evolution of a {\it system} is calculated up to second order in their coupling. In so doing, we are assuming that the coupling between the quantum system, i.e., the quantum spins, and the electronic reservoir is small, so that the reservoirs stay in its stationary state at thermal equilibrium, producing a {\em force} fluctuating about a zero average value with a short correlation time $\tau_c$, much shorter than the typical time scales of the spin system~\cite{Cohen_Grynberg_book_1998}.
  In our case, the {\it system} is the quantized spin while the {\it reservoir} is given by the conduction electrons and the coupling takes the form of the Kondo exchange \cite{Delgado_Palacios_prl_2010,Delgado_Rossier_prb_2010}.

The Bloch-Redfield theory yields the dynamical equation for the reduced density matrix~\cite{Cohen_Grynberg_book_1998}:
\beqa
\frac{d \hat\sigma_{nn'}(t)}{dt}  =-i\omega_{nn'}\hat\sigma_{nn'}(t)+\sum_{mm'}{\cal R}_{nn',mm'}\hat\sigma_{mm'}(t)
\label{master}
\eeqa
 with ${\cal R}_{nn',mm'}$ the Bloch-Redfield tensor, which  is  a quadratic functional of the system-reservoir coupling and is responsible of the dissipative effects.  Here we have employed the {\em secular} 
approximation for the master equation, which basically consist in neglecting the coupling between coherences with very different Bohr frequencies~\cite{Cohen_Grynberg_book_1998}. 
  Importantly, in a time scale much longer than the decoherence time $T_2$, the dynamics of the 
 diagonal terms, $ p_n\equiv \hat\sigma_{nn}$, which describe the occupations of the  $|n\rangle$
eigenstates of ${\cal H}_S$, 
and off-diagonal terms, $\hat \sigma_{nn'}$, that describe coherences between levels $n$ and $n'$, are decoupled.   The equation for the occupations yields the standard master equation:
\beqa
\frac{d p_n(t)}{dt}  =+\sum_{m} \Gamma_{m\rightarrow n } p_m - p_n\sum_{m} \Gamma_{n\rightarrow m}
\label{master2}
\eeqa
 where the transition rates from state $m$ to $n$, denoted by $\Gamma_{m\rightarrow n}$, are related to 
${\cal R}_{nn,mm}=-\Gamma_{m\rightarrow n}$ for $n\ne m$ and ${\cal R}_{nn,nn}=-\sum_{m\ne n}\Gamma_{n\rightarrow m}$.

The Bloch-Redfield equation (\ref{master}) couples coherences $\hat\sigma_{nm}$ to $\hat\sigma_{n'm'}$~\cite{Cohen_Grynberg_book_1998}
only if the transition energy $\Delta_{nm}$ is degenerate with $\Delta_{n'm'}$.  In the two specific cases considered here,  there is no coupling between different degenerate coherences, so that their equation of motion reads:
\beqa
\frac{d \hat\sigma_{nn'}(t)}{dt}  =-i\omega_{nn'}\hat\sigma_{nn'}(t) -i  \left( \delta \omega_{nn'} 
+ i\gamma_{nn'}\right)\hat\sigma_{nn'}(t).
\label{master3}
\eeqa
Thus,  in an isolated system $(R_{nm,n'm'}=0)$,  Eq. (\ref{master3}) describes the evolution of the coherence between two levels $n$ and $n'$ as an
 oscillating function with angular frequency 
$\omega_{n,n'}=(E_{n}-E_{n'})/\hbar$.  The coupling to the reservoir has then two effects on the evolution of the coherences, given by the real and imaginary  parts of the Redfield tensor,   ${\cal R}_{nm,nm}\equiv -\gamma_{nm}-i\delta\omega_{nm}$~\cite{Cohen_Grynberg_book_1998}.
First, over a time scale $T_2 (nm)=1/\gamma_{nm}$  it induces
a decay  of the amplitude of the oscillation, known as decoherence.
 The decoherence rate,
$\gamma_{nm}$, 
has  both an inelastic population scattering term,
and the pure dephasing rate, $\gamma_{nm}^{ad}$, which does not require population scattering~\cite{Cohen_Grynberg_book_1998}. 
 Second, it modifies the oscillation frequency $\omega_{nn'}$, i.e. it induces a shift of the transition energy.  Therefore, decoherence and energy shift are the real and imaginary part of the same  Bloch-Redfield tensor and are both consequences of the dissipative coupling of a quantum system to a fluctuating reservoir.

In the following we  provide explicit expression for the scattering rates, decoherence rates and transition energy shifts  for the case of a quantized spin exchanged coupled to an electron gas.

\subsection{Transition rates}
We consider the simple case where the quantized spin is coupled to a single electronic reservoir that is not spin polarized.  The rates in the simpler case read:
\beqa
\Gamma_{n\rightarrow m}^{SS}=\frac{1}{\hbar}\sum_{a=x,y,z} \left|\langle n|S_a|m\rangle   \right|^2 {\rm Im}\left[{\cal I}(\Delta_{nm})\right],
\label{ratesg}
\eeqa
where the $\Delta_{nm}=E_n-E_{m}$ and ${\rm Im}[{\cal I}]$ denotes the imaginary part of:
\begin{equation}
  {\cal I}(\Delta)=
2 \int d\epsilon d\epsilon' \rho(\epsilon)\rho(\epsilon') \frac{\left|J_{k(\epsilon)k(\epsilon')}^{SS}\right|^2f(\epsilon)\left(1-f(\epsilon')\right) }{\epsilon-\epsilon'+\Delta + i0^+}  ,
  \label{integralPP}
\end{equation}
with $f(\epsilon)$ the Fermi function and $\rho(\epsilon)$ the electronic density of states.  The cases of a spin polarized reservoir and coupling to two reservoirs yield analogous expressions and have been discussed elsewhere \cite{Delgado_Palacios_prl_2010, Delgado_Rossier_prb_2010}.

We can give a closed analytical expression for the transition rates, in terms of the spin matrix elements, the Kondo exchange, and the density of states of the conduction electrons at the Fermi energy if we   assume  a flat density of states $\rho=\frac{1}{W}$ with an energy bandwidth $W$,  and we neglect the momentum dependence of  Kondo exchange  $J_{kk'}^{SS}=J/L^d$.  With those assumptions, and taking into account the distribution relation $\frac{1}{x+i\epsilon}=-i\pi \delta(x)+ {\cal P}\frac{1}{x}$
where ${\cal P}$ stands for the Cauchy principal part,  Eqs. (\ref{ratesg}-\ref{integralPP})  take the form
\begin{equation}
 \Gamma_{n\rightarrow m}^{SS}=\frac{\pi}{2\hbar} \left(\rho J\right)^2 {\cal G}(\Delta_{nm})
 \sum_{a}\left|\langle n|S_a|m\rangle   \right|^2,
\label{ratesS}
\end{equation}
where 
\begin{equation}
{\cal G}(\Delta)\equiv\int d\epsilon f(\epsilon) \left(1-f(\epsilon+\Delta)\right)=\frac{\Delta}{1-e^{-\beta\Delta}}
\end{equation}
and $\beta^{-1}=k_BT$.
In the limit $\Delta\gg k_BT$ we can approximate ${\cal G}(\Delta)\approx \Delta$ which implies that the spin relaxation  rate from $n$ to $m$ is proportional to the energy released in the transition.  In contrast, the same equation yields that uphill transitions are thermally suppressed by a Boltzmann factor.   In the elastic case ($\Delta=0$), we have $ {\cal G}(0)=k_bT$, which is the well known Korringa results for spin relaxation\cite{Korringa_phys_1950}. Thus, the spin relaxation rate is governed by the dimensionless parameter $\rho J$, the spin selection rules contained in $\sum_{a}\left|\langle n|S_a|m\rangle   \right|^2$ and the phase-space  for scattering described by $ {\cal G}(\Delta)$.

\subsection{Decoherence rates}
The adiabatic decoherence rate for a  coherence between two degenerate states $n$ and $m$ of a quantized spin exchange coupled to a
 spin unpolarized electron gas, using the same approximations of momentum-independent  $J$ and flat density of states $\rho$, 
 reads: 
\beqa
\gamma_{nm}^{ad}&=&\frac{\pi}{4\hbar}\left(\rho J\right)^2k_B T
\crcr
&\times&
\sum_{\sigma\sigma'}\left| \sum_{a=x,y,z}\left(\tau_{\sigma\sigma'}^a\langle n|S_a|n\rangle -\tau_{\sigma'\sigma}^a\langle m|S_a|m\rangle \right)\right|^2, 
\label{decohr}
\eeqa
where $\tau^a$ is the $a$-Pauli matrix.  Notice that this adiabatic decoherence occurs via elastic scattering.  The adiabatic rate scales linearly with  $k_BT $ because elastic scattering  events require that the conduction electron system has an initial occupied state degenerate with a final empty state.
As in the case of the scattering rates,   decoherence rates  $1/T_2$  is also  proportional to $(\rho J)^2$. 

\subsection{Energy shifts\label{Eshifts}}
We now write down the expression for the shift of a transition energy $\tilde \Delta_{nm}\equiv \Delta_{nm}+(\delta E_n-\delta E_m)$ between two non-degenerate eigenstates $n$ and $m$ of the quantized spin Hamiltonian due to their Kondo coupling to an electron gas:
\begin{equation}
  \delta E_{n}=
\sum_{a,n'}\left|\langle n|S_a|n'\rangle   \right|^2 {\it Re}\left[{\cal I}(\Delta_{nn'})\right]
  \label{dEn}
\end{equation}
where  Re$[x]$ corresponds to the real part. Notice that the shift of a level $n$ has contributions coming from spin-matrix elements connecting to all states $n'$. 
  At the same level of approximation used to derive  Eqs. (\ref{ratesS}) and (\ref{decohr}),
 we obtain an expression for the shift of an energy level $n$~\cite{Oberg_Calvo_natnano_2013,Cohen_Grynberg_book_1998} 
\beqa
  {\rm Re}\left[{\cal I}(\Delta)\right]
\approx 
\frac{\left(\rho J\right)^2}{2}\int_{-W}^{W} d\epsilon f(\epsilon){\cal F}(\epsilon),
  \label{integralS}
\eeqa
where 
\beqa
{\cal F}(\epsilon)=
-\ln\left(\frac{2\pi k_B T}{-\epsilon-\Delta+W}\right)
-
{\rm Re}\Psi\left[\frac{1}{2}-i\frac{\epsilon+\Delta}{2\pi k_B T}\right],
\eeqa
with $\Psi(x)$ the digamma function. This expression is valid provided $W\gg k_BT$~\cite{Bloomfield_Philip_pr_1967}, which for metals stands even at room temperature.
In the regime we are interested in, where the splittings $\Delta_{nm}$ are much smaller than the 
bandwidth,  Eq. (\ref{integralS}) can be approximated by its leading order term in $W/(k_BT)$:
\begin{equation}
  {\rm Re}[{\cal I}(\Delta)]\approx {\rm Re}[{\cal I}(0)]-\frac{\Delta }{4} \left(\rho J\right)^2 \ln \frac{2W}{\pi k_B T},
  \label{integapp}
\end{equation}
where $ {\cal I}(0)$ is a shift common to all the energy levels, and therefore, not observable. Thus, if we consider the excitation energy of an anisotropic spin system, the renormalized energy difference $\tilde \Delta_ {nm}$ will be given to leading order in  $W/(k_BT)$ as
\beqa
\tilde \Delta_{nm}&\approx&\Delta_{nm}-\frac{\left(\rho J\right)^ 2}{4}\ln\frac{2W}{\pi k_B T}
\crcr
&\times &
\sum_{n',a}
\left[
\left| \langle n|S_a|n'\rangle\right|^2\Delta_{nn'}-\left| \langle m|S_a|n'\rangle\right|^2\Delta_{mn'}\right].
\label{shiftG}
\eeqa

Thus, the renormalization of the transition energy $\Delta_{nm}$ is the difference in the shifts of the levels $n$ and $m$, both described with Eq. (\ref{dEn}), see Fig. \ref{fig2}a.

\subsubsection{Second order correction of the $g$ factor for $S=1/2$}

We first apply Eq. (\ref{shiftG}) to the case of a spin $S=1/2$ interacting with a magnetic field. Including the term linear in $\rho J$ from Eq. (\ref{Knight}), we obtain the following expression up to order $(\rho J)^2$: 
\begin{equation}
\tilde \Delta_{\uparrow,\downarrow}= \Delta^0_{\uparrow,\downarrow} \left(1- \frac{g_e}{2}\rho J - \frac{1}{4}(\rho J)^2 \ln\frac{2W}{\pi k_B T}\right)
\end{equation}
In the derivation of the contribution quadratic in ($\rho J$) to this equation we have ignored the small difference in density of states for different spin orientations,  that plays a role in the derivation of the rates \cite{Delgado_Rossier_prb_2010}. By so doing, though, we 
recover the same expression that was derived using the Kadanoff-Baym Green function method by Langreth and Wilkins~\cite{Langreth_Wilkins_prb_1972}.

\subsubsection{Shifts for half-integer spins}
\begin{figure}[t]
  \begin{center}
  \includegraphics[width=1.\linewidth,angle=0]{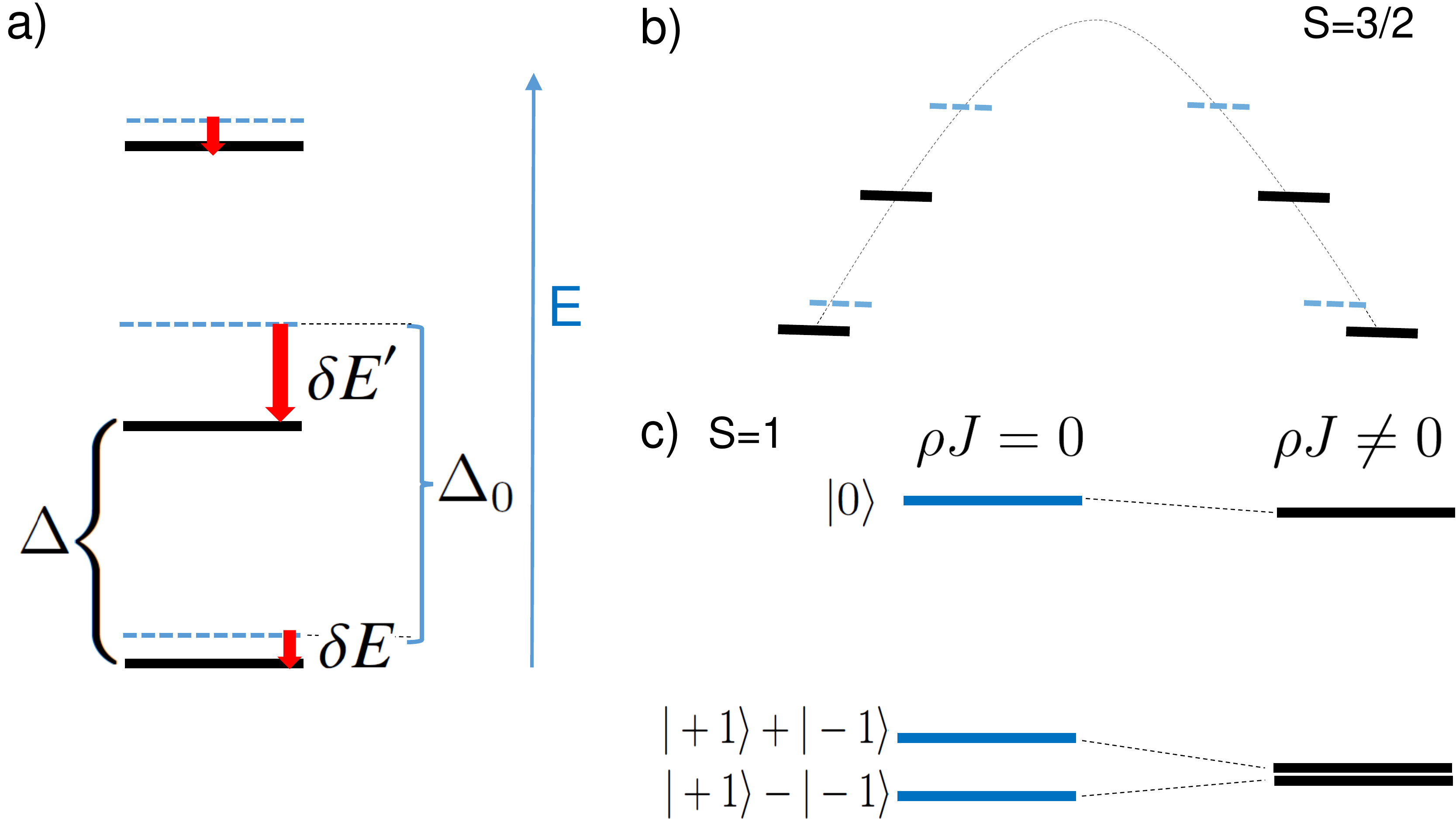} 
  \end{center}
  \caption{a) Sketch of the renormalization of the energy levels of an anisotropic spin system. b) Modification of the spin levels of a $S=3/2$ spin system with easy axis anisotropy ($D<0$).
The finite exchange with the substrate electrons induces a reduction of the excitation energy $\Delta_0=E_{3/2}-E_{1/2}$. For finite $\rho J$, each energy level $E_n$ is shifted with a displacement proportional to $\left( \rho J\right)^2$, leading to an excitation energy $\Delta\le \Delta_0$. c) Quenching of the zero field splitting of a $S=1$ spin system with the exchange coupling $J$.
    \label{fig2}
  }
\end{figure}

We now compute the shift of the spin transition for a system with finite spin excitations at zero field, namely,
 a $S=3/2$ spin system described with Hamiltonian (\ref{spin-hamil}), whose energy levels are shown in figure 2(b).. This, for instance,  describes a  
Cobalt adatom on Cu$_2$N~\cite{Otte_Ternes_natphys_2008,Oberg_Calvo_natnano_2013}  . In that case there is a single transition at zero-field and, for $|D|\gg E$ in Eq. (\ref{spin-hamil}), the only excitation energy is given by~\cite{Oberg_Calvo_natnano_2013}
\begin{equation}
 \tilde \Delta\equiv \tilde E_{3/2}-\tilde E_{1/2} 
  \approx 
  \Delta_0 \left(1- (\rho J)^2  \frac{\Lambda}{8}
  \ln\frac{2W}{\pi k_BT}\right), 
  \label{delta1}
\end{equation}
where $\Delta_0=(E_{3/2}-E_{1/2})$ and 
$\Lambda=\sum_a|\langle 1/2|S_a|3/2\rangle|^2$ (see Fig. \ref{fig2}b). In particular, we have that to lowest order in $D/E$, $\Lambda\approx 3/2(1+E^2/D^2)$.  The renormalization of the excitation energy $\tilde\Delta$ has been recently measured\cite{Oberg_Calvo_natnano_2013} for Co adatoms on a Cu$_2$N substrate, where the presence of a Kondo peak in the conductance permits calibrating the intensity of the exchange coupling. There, the variations of the Kondo coupling throughout the Cu$_2$N islands led to variations of the magnetic anisotropy by a factor 2, from $\tilde\Delta\approx 10$ meV and axial and transversal magnetic anisotropies $D\approx 3.5$ meV and $E\approx 2$ meV, respectively, to a reduced excitation energy $\tilde \Delta\approx 5$ meV and  $D\approx 5$ meV and $E\approx 0$ when $J\to 0$.

\subsubsection{Shifts for integer spins}
We now apply apply the theory to study the renormalization of the zero-field splitting $\Delta_0$ due to the quantum tunnelling of magnetization present in integer spins when $E\ne 0$. 
 For simplicity, we consider the example of a spin $S=1$ with easy axis anisotropy ($D<0$) (see figure 2(c))..   The eigenvalues of  this system are $E_{\alpha}=-D-E$, $E_{\beta}=-D+E$ and $E_0=0$.   We assume $D<0<E<|D|$ so that $E_{\alpha}<E_{\beta}<E_0$.  Using the general expressions for the shifts, Eq. (\ref{shiftG}), one gets that
\beqa
\label{deltazfs}
\tilde \Delta_{\beta,\alpha}=\Delta_{\beta,\alpha}\left[1-\frac{3}{2}\left(\rho J\right)^2 \ln\frac{2W}{\pi k_B T}\right].
\eeqa
where $\Delta_{\beta,\alpha}=2E$ is the bare energy splitting and we have taking into account that the contribution coming from the matrix elements within the $\alpha,\beta$ doublet is twice the one coming from their  coupling to the $0$ level, see Eq. (\ref{shiftG}).

Hence, in the case of integer spins, the Kondo coupling is the responsible for the renormalization of the quantum tunnelling splitting, enabling in this way the decoherence between the two quantum mechanical superposition states and being ultimately responsible of the emergence of the classical magnetic states~\cite{Delgado_Loth_prep}. 

\section{Discussion and conclusions}
We have reviewed the effect of the Kondo exchange  on the spin dynamics of a  quantized spin in the weak coupling limit where perturbation theory works. We have discussed four different effects: 
\begin{enumerate}
\item The renormalization of the $g$ factor, related to the so called Knight shift in nuclear physics (Eq. (\ref{Knight})). 
This might account in part for the 
observed anomalously large $g$ factor of single Fe atoms adsorbed on a Ag(111) \cite{Chilian_Khajetoorians_prb_2011}.

\item  The finite lifetime of excited spin states, which gives rise to a relaxation rate of excited spin states (Eq. (\ref{ratesS})).  The linear relation between the rate and the transition energy might account for the observations of for Fe on Ag(111) \cite{Chilian_Khajetoorians_prb_2011} as well as for Fe on Cu(111) \cite{Khajetoorians_Lounis_prl_2011}.

\item The decoherence of degenerate eigenstates of the spin Hamiltonian [Eq. (\ref{decohr})].  In the case of semi-integer spins with $D<0$ this mechanism 
ensures that coherence between the two ground states with opposite $S_z=\pm S$ disappears on a timescale of picoseconds~\cite{Delgado_Rossier_prl_2012},  preventing the formation of Schr\" odinger-cat like states with null magnetization~\cite{Delgado_Loth_prep}. 

\item The renormalization of the excitation energy $\Delta_{n,m}$ [Eq. (\ref{shiftG})] that accounts for the recent observation for Co atoms on Cu$_2$N/Cu(001) \cite{Oberg_Calvo_natnano_2013}. 

\end{enumerate}
To this list of physical phenomena, controlled by  $\rho J$,  that involve  quantized spins exchanged coupled to an electron gas, one should of course add the Kondo effect and, when two or more quantized spins are coupled to the same electron gas, the indirect exchange interaction $RKKY$.

In this work we have reviewed the case of a quantized spin couple to a single spin unpolarized electrode (except during the discussion of linear order effects).   It is straightforward  to extend  the theory  in two directions. First, one can include a second electrode at a different chemical potential to model inelastic electron tunnel spectroscopy \cite{Rossier_prl_2009}. In this framework, the inelastic current is proportional to $\rho_T \rho_S J_{TS}^2$, where $\rho_{T,S}$ are the density of states at the Fermi energy of tip and surface and $J_{TS}$ is the Kondo exchange coupling for processes in which the electron tunnels between tip and sample. 

The second extension is to consider a spin-polarized electrode.  This permits modelling \cite{Delgado_Palacios_prl_2010} two additional effects that have been reported in the literature of magnetic adatoms\cite{Loth_Bergmann_natphys_2010,Khajetoorians_Lounis_prl_2011,Khajetoorians_Wiebe_science_2011}.  For a spin polarized STM tip, the linear conductance depends on the relative orientation of tip and magnetic adatom magnetizations.  The spin contrast is linearly proportional to $J_{TS}$.   In addition,  spin-polarized currents can control the orientation of the magnetic adatom\cite{Loth_Bergmann_natphys_2010}. 

As a final remark, we have to note that whereas the theory presented here looks different from other theory work addressing the same problem \cite{Lorente_Gauyacq_prl_2009,Khajetoorians_Lounis_prl_2011}, the underlying physical phenomena are the same.  The main advantages of the approach reviewed here are the following. First,  it permits   connecting the physical phenomena  observed in magnetic adatoms with well established effects, proposed in other contexts, such as the Knight shift~\cite{Knight_pr_1949} or the Korringa spin relaxation~\cite{Korringa_phys_1950}.  Second,  it  identifies a single dimensionless quantity, $\rho J$ as the essential parameter that controls the magnitude of the influence of Kondo exchange on the spin dynamics of magnetic adatoms.  Third, it connects with a vast body of literature using the Kondo model \cite{Wolf_Loose_physlett_1969,Langreth_Wilkins_prb_1972,Hewson_book_1997}.

\section*{Acknowledgements}

We acknowledge J. C. Oberg, M. R. Calvo, D. Jacob, D. Serrate and M. Moro-Lagares for fruitful discussions.
This work was supported by the Engineering and Physical Sciences Research Council UK (EP/D063604/1 and EP/H002367/1); the Leverhulme Trust (RPG-2012-754); Ministry of Science and Education Spain (FIS2010-21883-C02-01, MAT2010-19236, CONSOLIDER CSD2007-0010 and Programa de Movilidad Postdoctoral); European Commission FP7 Programme (PER-GA-2009-251791) and GV grant Prometeo 2012-11.

\section*{References}











\end{document}